# Automatization of building IT projects using composite consistency rules

*Stanisław Jerzy Niepostyn*
School of Computer Science & Technologies,
University of Economics and Human Sciences
in Warsaw, Okopowa str. 59, 01-043 Warsaw,
Poland
j.niepostyn@vizja.pl

*Wiktor Bohdan Daszczuk*
Institute of Computer Science, Warsaw
University of Technology, Nowowiejska str.
15/19, 00-665 Warsaw, Poland
wiktor.daszczuk@pw.edu.pl

**Abstract**

Unified Modeling Language (UML) is widely used for modeling IT systems but lacks formal rules to ensure consistency across diagrams. This often leads to inconsistencies when shared elements are interpreted differently. To address this, architects use consistency rules that derive elements in target diagrams from more abstract source diagrams. However, these rules are often written in natural language and applied at the element level, making them difficult to reuse or integrate with modeling tools.

This paper introduces composite consistency rules—higher-level patterns that combine simple rules into more intuitive, reusable structures. These rules reflect architects design practices and support systematic, error-resistant model development. Implemented as JScript scripts in Sparx Enterprise Architect, they improve automation, reduce redundancy, and accelerate design. Composite rules enhance the consistency and completeness of UML architectures and can be reused across projects. They also support pattern-driven modeling and open possibilities for AI-assisted architecture generation and code integration.

**Keywords:** UML diagrams, software architecture, consistency rules, composite consistency rules.

## 1. Introduction

The Unified Modeling Language [1] is a de facto standard for modeling software architecture of IT systems. This standard defines syntax and semantic rules for single diagrams. There are no rules defined between several UML diagrams. It leads to some uncertainty and can result in lacking order in the process of constructing software architecture, which can cause inconsistencies in the software architecture. The problem of model inconsistency was proposed by Spanoudakis and Zisman in 2001 [2]. They noticed that inconsistencies arise between models that describe the same system from different points of view and use common elements. Such a common element can be interpreted in various ways across models, so differences in its interpretation may lead to inconsistencies in the overall architecture. In order to deal with model inconsistency, IT architects use so-called consistency rules to create consistent and complete software architecture. In fact, so-called consistency rules are actually rewrite rules that generate elements of a target diagram on the base of elements occurring in higher level diagrams, preserving the consistency of the entire project. Such consistency rules have been proposed in the software architecture verification process since the publication of the first version of the UML standard. Furthermore, the proposed consistency rules are mainly described in the natural language. Consistency rules are usually used in software architecture in this way [3][4][5][6][7] and are usually not related to the modeling tool in any way. It frequently results in overlapping or duplicating rules within different papers. Hence, it is not easy to identify a set of unique rules and find their similarities.

In our previous paper [8], we defined the consistency and completeness of software architecture using entropy-based measures. The e-CMDA (Enhanced Consistent Model Driven Architecture) algorithm proposed a specific series of diagrams linked by consistency rules in fourth views: context, business, system, and development. Each of these views consists of two layers, which in turn consist of one to three practical diagrams based on UML diagrams. Adjacent diagrams (in the sense of the software architecture model presented in Table 1) are related to each other by means of consistency rules described in a formal using symbols denoting individual elements, available at [9]. Table 1 presents the successive practical diagrams created in the relevant views and layers of the e-CMDA software architecture

**Table 1.** The e-CMDA software architecture model.

| View | Layer | Project Diagram | UML Diagram | Mark |
|---|---|---|---|---|
| **Context** | Context | Context Diagram | Activity | X |
| | Process Decomposition | Business Use Case<br>Process Decomposition Diagram | Use Case<br>Activity | B<br>R |
| **Business** | Business Process | Business Use Case Realization Diagram | Activity | A |
| | Logic | System Use Case Diagram<br>Business Class Diagram<br>Business State Machine Diagram | Use Case<br>Class<br>State | U<br>C<br>S |
| **System** | User | System Use Case Realization Diagram | Activity | Z |
| | Internal | Implementation Use Case Diagram<br>System Class Diagram<br>System State Machine Diagram | Use Case<br>Class<br>State | Y<br>J<br>T |
| **Development** | Sequence | Implementation Use Case Realization Diagram | Sequence | Q |
| | Component | Component Diagram | Component | M |

This paper proposes new, higher level consistency rules based on the previously published list of consistency rules [9]. These newly proposed rules, which we call *composite rules*, are mixtures of previously published simple rules. The implementation of these composite consistency rules in the Sparx Enterprise Architect tool (in the form of JScript scripts) has proven that building a software architecture using the proposed new composite consistency rules significantly automates the software architecture design and shortens the design time, which undoubtedly affects the faster start of implementation and deployment of the IT system. The mechanism of developing new composite consistency rules is an important aspect of software architecture design, based on which one can consider the concept of using pattern-driven software architecture fragments in building the target software architecture, and even use artificial intelligence methods both in increasing the speed of building the software architecture, and even incorporating verified source codes into building IT systems.

Composite consistency rules give several benefits when compared to simple ones:
- Composite consistency rules are more intuitive and closer to the designers experience in software architecture design.
- Composite consistency rules are the patterns for systematic building of software architecture composed of UML diagrams.
- The application of composite consistency rules leads to the consistency and completeness of software architecture.
- Composite consistency rules can be reused in many IT projects.
- Using composite consistency rules to generate UML diagrams speeds up the whole process of designing an IT system and eliminates sources of errors.
- Using composite consistency rules to generate UML diagrams greatly increases the automation of the software architecture building process, which significantly reduces errors associated with using the UML standard.

- Composite consistency rules allow for obtaining an IT system design with greater consistency and completeness, with a high level of confidence in the correctness of the obtained software architecture
- Using pattern-driven composite consistency rules to generate UML diagrams further automates the software architecture building process and further reduces the time spent building software.
- The software architecture construction method is available to everyone, thanks to the development of ready-made scripts for Sparx Enterprise Architect, available in folder JScript at [9].

Section 2 describes the articles describing the area indicated in this article, and Section 3 describes the main part of the article - composite consistency rules. Section 4 gives an example of the application of the proposed consistency rules, and the article is summarized in Section 5.

## 2. Related work

Software models can present the system while considering different points of view, different levels of abstraction and granularity, different notations, etc. They may represent the viewpoints and goals of different stakeholders. Usually, some inconsistencies between models arise [2]. Inconsistencies in models may lead to design problems. Obviously, the earlier the problems are detected in the system design lifecycle, the lower the cost of fixing them.

One of the ways to deal with inconsistency in software models is to verify them using consistency rules. In general, consistency rules are transformations (mappings) between elements of different models. Egyed in [10] proposed the methods for fixing inconsistencies in UML diagrams. Those methods were regarding class, state, object, and sequence UML diagrams. Another approach to check consistency in activity diagrams was proposed by Jurack et al. in [11]. In this method, the consistency of the activity diagram was validated by checking whether all flow paths could be visited. Shinkawa, in his research [12], proposed to generate consistent UML diagrams from the activity diagram based on Colored Petri Net. A few rules for consistency between activity diagrams and use case diagrams were proposed by Ibrahim [13]. The majority of consistency rules from different authors have been gathered in Torre [3], but his set was specified in natural language.

Our research focuses on applying first-order consistency rules using the consistency rules published in [5] to construct entire software architectures. As in the aforementioned paper, we continue to represent consistency rules using the original notation based on regular expressions (UNCR) introduced (for example, + is one or more occurrences, | is a choice, ? is at most one occurrence). Our notation allows us to provide a much more precise and complete taxonomy of consistency rules and facilitates their comparison.

## 3. Composite Consistency Rules

The consistency rules published in the literature, which we will call *simple rules*, are intended to serve as rules for creating lower-level diagram elements from higher-level diagrams, ensuring the consistency of the project [9]. Generally, a simple consistency rule maps elements of a source diagram to a single element of a target diagram. The exceptions are the three consistency rules published earlier in [5] that map a use case scenario to a use case realization diagram (R2.5, R3.5, R4.6). The analysis of many IT projects by the authors of this paper shows that simple rules are used by designers, contrary to their intended use, for simple control of the correctness of the diagrams being created. However, when designing subsequent diagrams, architects and designers tend to use specific structures (compositions) in the source and target diagrams. We will show that the use of one composite consistency rule allows, based on the observation of specific structures in the source diagram, the creation of more than one element in the target diagram and linking them with appropriate relationships instead of introducing numerous simple rules, enabling the linking of elements of the target diagram.

In our approach, a *composite consistency rule* maps at least two linked elements of the source diagram to at least two linked elements of the target diagram. These connected elements from both the source and target diagrams model some aspect (some concern) of the software architecture, such as mapping several instances connected to a process to several actors using a specific use case (R1.22), or mapping the data flow through instances of one class to the life cycle of that class (R2.26a).

Below, we present in individual subsections, describing sample mapping groups, and the application of the proposed composite consistency rules. In the first subsection, for the contextual perspective, an example of using simple rules to build subsequent diagrams is described, followed by a composite consistency rule based on simple consistency rules. The following sections explain examples of the proposed composite consistency rules and, finally, another type of these rules, which we call pattern-driven composite consistency rules. We describe exemplary rules, the complete list of them, and the implementation in Enterprise Architect scripts can be found in [9].

### 3.1. Composite consistency rules from context diagram to process decomposition diagram

This section describes new composite consistency rules that can be used to build a process decomposition diagram based on a context diagram. First, an example of using simple rules to build a process decomposition diagram is described. Then, the construction of a process decomposition diagram using the proposed composite consistency rules is presented.

The notation in a consistency rule is as follows: uppercase letters denote the source and target diagrams (on the left and on the right, respectively), and lowercase letters denote the diagram elements that the rule applies to (e-event, v-activity/action, c-class, i-instance, z-association, a-actor, u-use case, n-control node, r-region, s-state invariant, h-operation, l-lifeline, m-message, q-component, for the complete list please refer to [9]). In addition, elements can be provided with parameters in parentheses, or stereotypes in double angle quotation marks. In addition, rules take the form of a regular expression, where + denotes one or more occurrences of the element, and | denotes a choice. The parentheses in a regular expression have an obvious meaning, unless they enclose a parameter, which is evident from the context. States are given in square brackets.

Figure 1 shows previously published rules [9] that were used to create the appropriate structure in the process decomposition diagram based on the context diagram. It should be noted that to create a subprocess with its input object and output object, 5 simple rules placed in four elements of the context diagram should be used. First, on the "Request" event in the context diagram, the XeRi rule should be entered (based on the "Request" event from the context diagram, an object of the "Request" class should be created in the process decomposition diagram). Then, on the same element, the XevRv rule should be entered (based on the "Request" event and the main "Office" process from the context diagram, a subprocess named "1.Request_service" located in the rule parameter should be created in the process decomposition diagram). On the next object of class "Decision" on the context diagram, enter the rule XiRi (based on the object of class "Decision" from the context diagram, create an output object of class "Decision" on the process decomposition diagram). On the control flow connecting the "Request" event with the main process "Office" on the context diagram, enter the rule Xz«control»Rz«data» (based on the control flow from the "Request" event to the main process "Office" from the context diagram, create a data flow on the process decomposition diagram connecting the input object of class "Request" with the subprocess named "1.Request_service" in the rule parameter). In the data flow connecting the main process "Office" with the object of class "Decision" on the context diagram, the rule Xz«data»Rz«data» should be introduced (based on the data flow from the main process "Office" to the object of class "Decision" from the context diagram, creating a data flow connecting the subprocess named "1. Request service" with the output object of class "Decision" on the process decomposition diagram).

This method of creating a structure on the process decomposition diagram based on elements from the context diagram seems quite laborious and time-consuming. Hence, it is not surprising that this method of building architecture has not gained popularity and recognition (however, it should be noted that it could be useful and provide consistency).

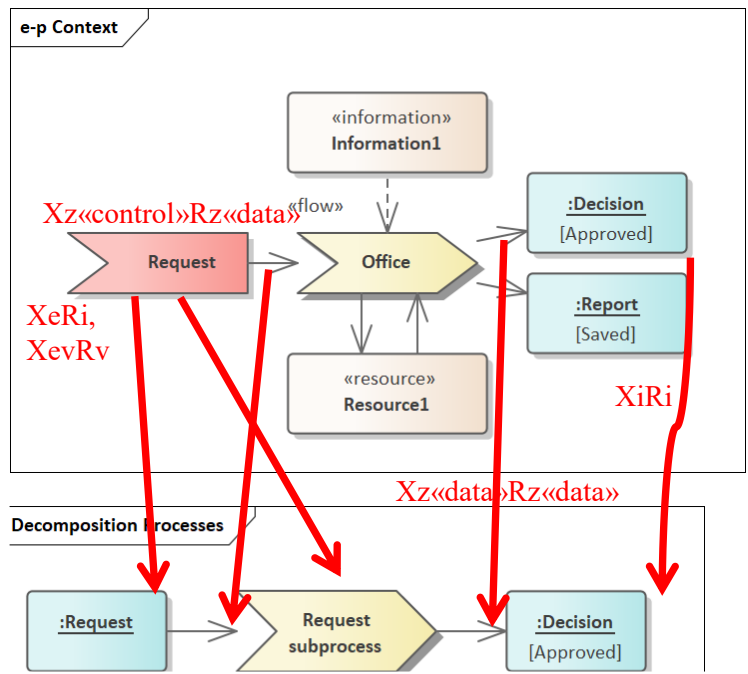

**Fig. 1.** Application of a set of simple consistency rules in the construction of a process decomposition diagram.

In Figure 2, however, the same construction of 5 simple rules can be obtained using a single composite consistency rule placed on one of the elements that participates in the rule used to build the target construction in the process decomposition diagram.

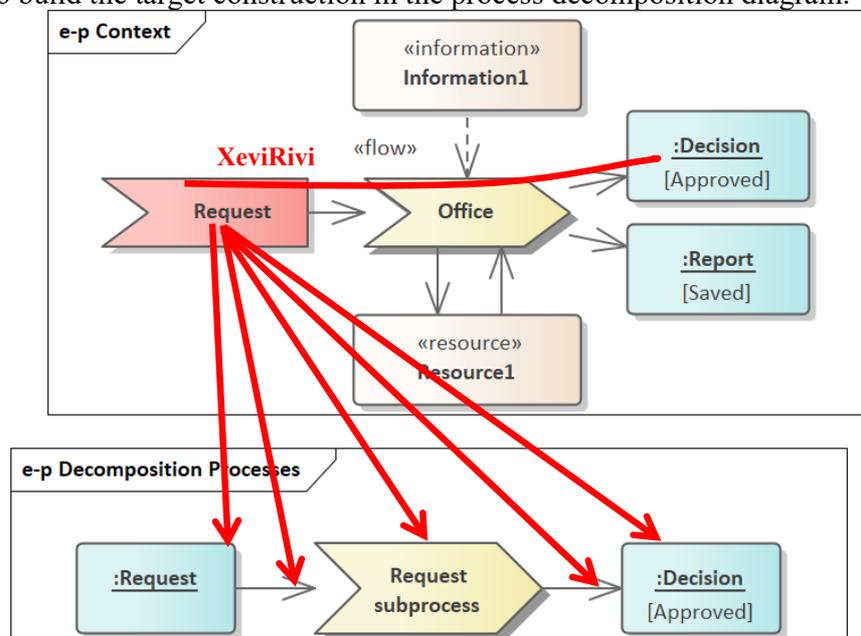

**Fig. 2.** Application of composite consistency rules in the construction of a process decomposition diagram.

This complex consistency rule can be denoted as XeviRivi, or more precisely as: Xevi(:Decision)Ri(:Request)v(1.Request_service)i(:Decision). The parameters of this rule are placed in parentheses. These parameters denote the names of the source elements from the context diagram and the names of the target elements that should be created in the process decomposition diagram. So, instead of entering five simple rules on four elements of the context diagram, it is enough to enter one complex consistency rule on one of the elements of the context diagram. The mechanism of complex consistency rules dramatically simplifies and reduces the time needed to create the appropriate construction in the target diagram. The description of consistency rule parameters will be described in detail in the next section.

Other composite consistency rules (from the group of rules available from the context diagram) that can support the design of a decomposition diagram are the decomposition of a main process into subprocesses by products and the decomposition of a main process into subprocesses by event and product into products and subprocesses (R1.5a and R1.5.c in [9]).

### 3.2. Composite consistency rules from process decomposition diagram to business use cases diagram

In addition to the published simple consistency rules [9], our design practice and cooperation with students in didactic classes allowed us to discover many additional, uncatalogued simple rules. They are the basis for creating a new group of composite rules. Namely, usually, after creating a process decomposition diagram, a business use case diagram is created. Creating this second diagram based on a process decomposition diagram is much easier than creating one based on a context diagram. Hence, the composite rules below are not based on any previously published consistency rules, because there were no such rules between a process decomposition diagram and a business use case diagram.

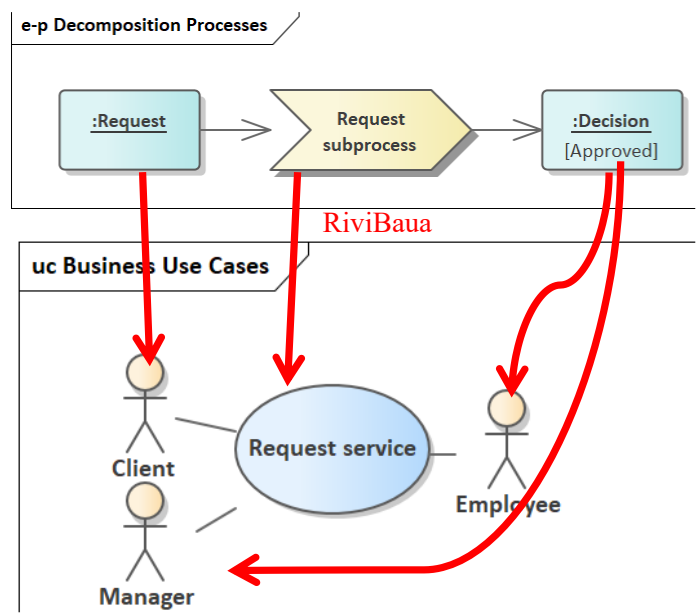

**Fig. 3.** Application of composite consistency rules in the construction of a business use cases diagram.

Figure 3 shows the composite rule (R1.20 in [9]) of the form Rev«subprocess»i«product»Baua (abbreviation: ReviBaua), which allows mapping of event and product to actors and mapping of subprocess to business use case. It is worth noting that only one composite consistency rule, placed on one element of the decomposition processes diagram, allows the creation of one use case and three actors associated with this use case.

### 3.3. Pattern-driven consistency rules for constructing a business use case realization diagram

An interesting group of composite consistency rules is the rules that enable the creation of individual elements in a business process diagram (business use case realization diagram) based on a business use case diagram or a process decomposition diagram.

An important rule in this group is the already published rule R2.5 (generating business use cases realization diagram from business use case scenario) in the form: Bu«scenarios»An«start»(v+|i+)+n«stop» (in short: BuAn(v+|i+)+n). This rule is not a simple rule, but is in fact a composite pattern-driven rule. This means that this rule allows us to create a target structure of interconnected elements by entering a specified construction pattern and adapting selected elements of that structure as if multiple composite consistency rules had been applied to create that target structure. After a minor modification, this rule takes the index R2.5a and the form: BauA(pv+)+. This rule allows us to create a control flow on the process diagram by creating a partition and activities placed in it. An example of the pattern is shown in the lower diagram of Figure 4. This pattern contains a process flow between anonymized actors and objects of the ObjectA class (the object at the process input) and objects of the ObjectZ class (the object at the process output). A composite rule driven by the pattern placed on the use case (R2.5a: BauApv({Office})) can specify the names of the actors, which will be converted into partition names, and the second composite rule driven by the pattern placed on the subprocess (R2.9a: RiviAv({Office})i) can specify the names of the object classes at the process input and at the process output. Therefore, in order to create the target diagram, we first need to execute a composite rule on the business use case diagram element (R2.5a – renaming partitions), and in the second step, we need to execute the composite rule (R2.9a - renaming instance classes) placed on the process decomposition diagram element. Figure 4 shows the mechanism of the cascade-like operation of these two new composite rules. Based on these pattern-driven composite rules, a business process diagram is created as discussed in Section 4.2.

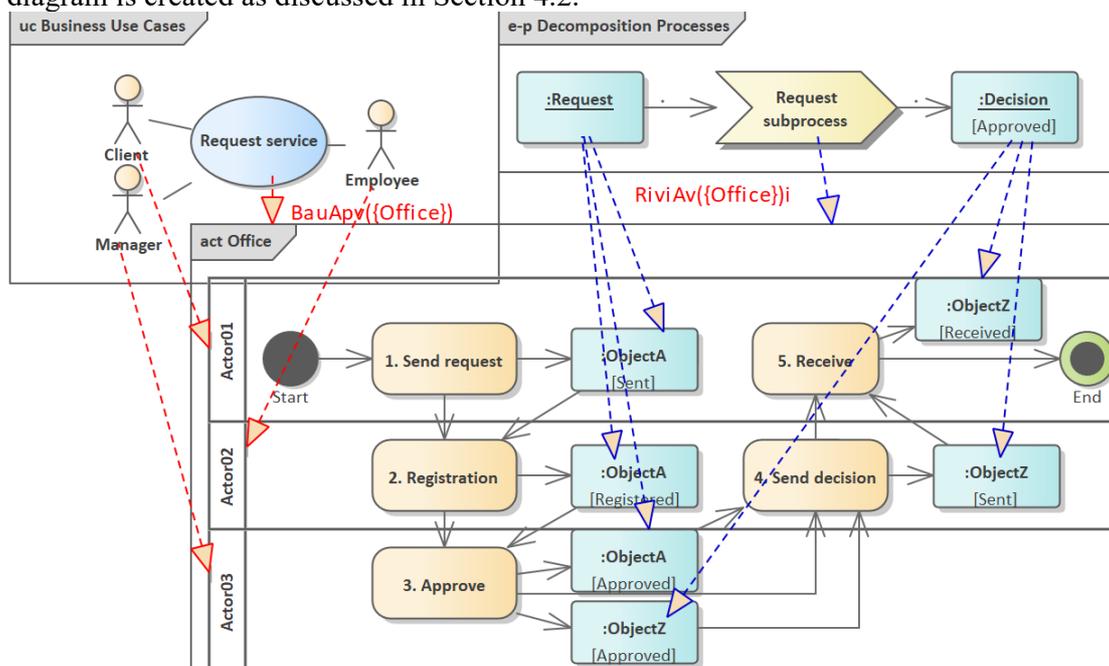

**Fig. 4.** Application of composite consistency rule in the construction of a process diagram.

At this point, it should be noted that the parameterized form of both composite consistency rules differs for various IT systems. This means that for each subsequent IT system, the business process diagram (Request processing in the office) may be completely different from the previously designed business processes. Hence, we are aware that the parameterized form of these rules requires appropriate time to analyze and design this business process individually, not only for a given IT project, but also for each of its business processes.

It is worth mentioning here that in the Enterprise Architect tool by Sparx there is a functionality that allows you to create an activity diagram based on previously described scenarios. However, even in this case, the designer must first describe the process scenario, in order to then obtain a business process diagram.

Hence, we are currently working simultaneously on the concept of selecting a specific business process from a set of previously developed business processes. When a user selects a specific business process pattern (for example, the previously mentioned business process pattern called processing a request in an office: Office), the business process diagram is automatically imported to Sparx Enterprise Architect, instead of (or in addition to) the parameterized composite consistency rules used. In the current state of the developed scripts for Sparx Enterprise Architect, it is enough to place a rule with the following parameter BauApv({Office}) or RiviAv({Office})i on the Business Use Case or subprocess element (the name of the business process pattern is placed in curly brackets). The "{Office}" parameter allows us to select a pattern, as in Figure 4, and create a process diagram called "Business Process Diagram" in a package called "Business Process Layer" as will be shown in Section 4.2.

In our approach, the *pattern* is a diagram of a business process (or a visualization of a specific scenario) implemented in a given organization in a way that we assume based on the execution of many IT projects. Currently, an example pattern is a business process called "Office", but it can also be a business process for a request for access to IT resources for a person outside the organization, or a business process for handling a request for reporting the location of a herd. Currently, available patterns are placed in folder Pattern at [9].

On the other hand, a pattern-driven composite consistency rule is a rule in which the parameter is the name of the business process pattern (or the name of the scenario visualization pattern). It means creating a target diagram (target structure on the diagram) according to the pattern with the given name. Therefore, instead of modeling a possible scenario of request processing in the office, the IT Architect can simply provide the name of the pattern in the composite consistency rule: "Office".

At the current stage of work, the concept of a composite pattern-driven consistency rule seems useful for people designing business processes in the Enterprise Architect tool and using scripts developed by us to build the software architecture of the IT system. In this concept, we will gradually enrich the current repository with previously detected business process patterns in already completed IT projects. It also seems that artificial intelligence mechanisms will be used, because even in similar business process flows there will certainly be different names of actions or even different names of instances occurring in the data flow.

To sum up this subsection, it should be noted that the proposed composite rules cannot be directly applied to the automation of the software architecture construction process without the designer intervention, but work is currently underway on the use of business process patterns, ready to replace or supplement these specific ones, i.e., pattern-driven parameterized composite consistency rules pattern. Moreover, it should be noted that currently, in the field of software architecture design, or business process design we have not found similar solutions.

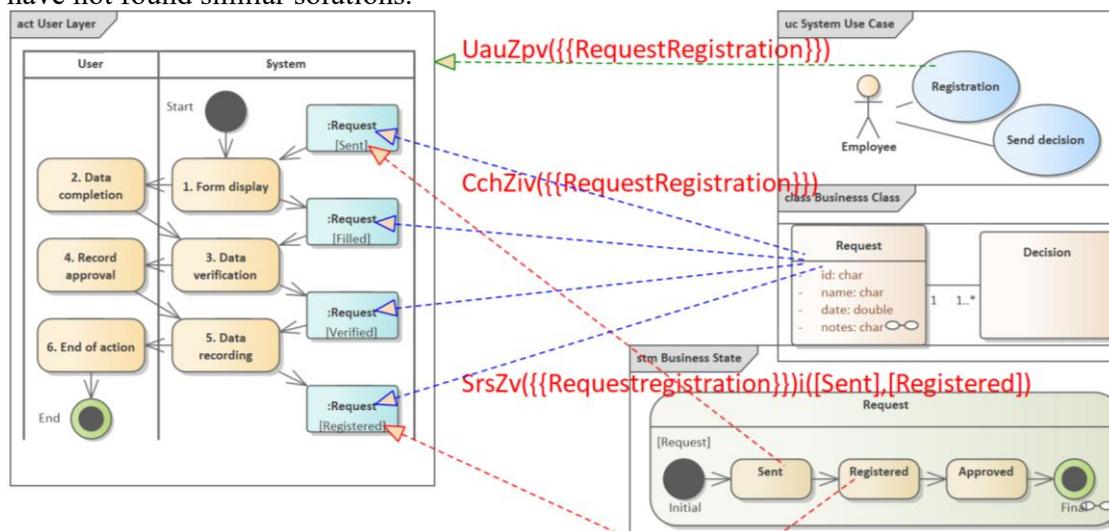

**Fig. 5.** Application of composite consistency rule in the construction of a system use case realization diagram.

On a similar principle, we also based the operation of pattern-driven composite consistency rules for visualizing system use case realization diagrams, as well as implementation use case realization diagrams (sequence diagrams). In these cases, the proposed new composite consistency rules also require prior analysis and design of the course of the relevant scenarios before they are introduced to Sparx Enterprise Architect. Instead of these composite consistency rules, we also proposed new pattern-driven composite consistency rules.

The first composite rule for mapping the system use case, and its actors to the activity control flow in the system use case realization diagram has the name: mapping the system use case and its actors to the activity control flow in the system use case realization diagram (index R3.5a) and the form: UauZ(pv+)+. Applying this rule with the parameter: UauZpv({{RequestRegistration}}) will create (or complete) the system use case realization diagram as in Figure 5. A similar effect can be obtained for the rule: data flow mapping taking into account the states of input objects and the states of output objects (R3.12b) of the form SrsZ(v+i+)+ when the parameter is substituted by: SrsZv({{RequestRegistration}})i([Sent],[Registered]) (the name of the system use case realization diagram pattern is placed in double curly brackets).

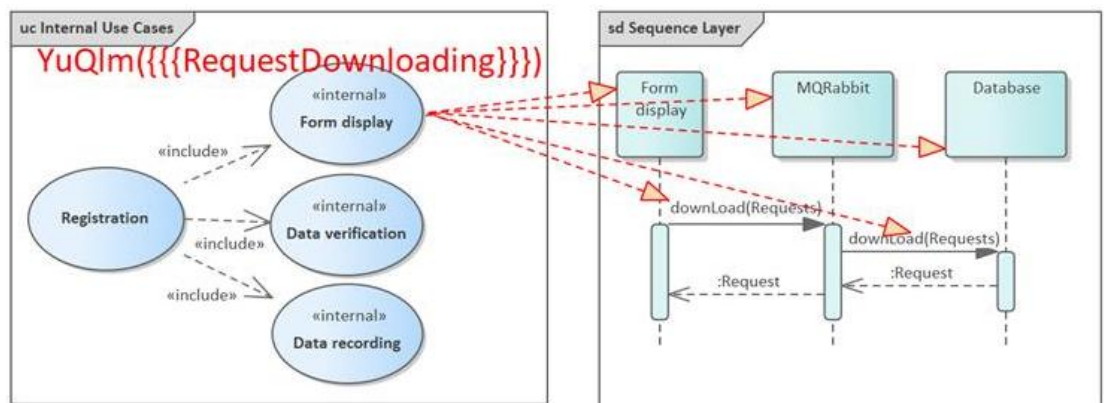

**Fig. 6.** Application of the pattern-driven composite consistency rule in the construction of an implementation use case realization diagram.

Another composite rule for mapping scenario steps to message transfer between lifelines in a sequence diagram: mapping scenario steps to message sending between lifelines elements (index R4.6a) of the form: YuQ(lm+)+. Applying this rule with the parameter: YuQlm({{RequestDownloading}}) will create (or complete) the implementation use case realization diagram (sequence diagram) as in Figure 6.

## 4. Example of Applying Composite Consistency Rules

This section presents an example of a modeled fragment of software architecture for a sample applications handling office.

### 4.1. Modifying target element names using consistency rule parameters

The consistency rule parameters play a fairly important role in constructing individual software architecture diagrams. Filling in the parameter will result in the creation of a target element with the name as in the parameter. This is simple in the case of target elements such as action, activity, event, actor, or use case. However, in the case of class instances, the issue is more complicated. This results from the fact that in addition to the instance name, it is also possible to specify the type of class from which this instance was created. In addition, the instance may have its own state, which increases the complexity of the mechanism for creating a target diagram element of the type "class instance". In this solution, it was assumed that the name of the target element is determined primarily by the name placed in the parameter. This name can take the form: name1[State1]:class1. This name may be correct, as long as the target element is an instance of the class. For other types of elements, the fragment: name1 is correct. However, if there is no value in the target element parameter, then it should be assumed that the class of this instance is the class of the source element, or the class of the source element instance, or the class with the name of the source event. It should also be noted that the consistency rule of the form $Xi^{STATE}Ri^{STATE}$ can actually be implemented only by the rule Xi([State])Ri([State]), i.e., by another rule, but with the state in the parameter name. Another important type of parameter is a business process pattern, a system use case realization diagram pattern, or a sequence diagram pattern (they are placed in single, double, or triple curly brackets, respectively). This notation is used in the next subsection.

### 4.2. Example of software architecture driven by composite consistency rules

Figure 7 presents a fragment of the software architecture that was created using only the new composite consistency rules proposed in this article.

In the context layer of the context perspective, the "Office" system model was placed. Then, in the event element named "Request", the XeviRivi rule was placed with the following parameters: Xevi(:Decision)Ri(:Request)v(1.Request_service)i(:Decision). This rule enabled the creation of a process decomposition diagram in the process decomposition layer. In this diagram, the RiviBaua rule was then placed in the Subprocess element with the following parameterized form: Rivi(:Decision)Ba(Client)ua(Employee,Manager). This rule, in turn, enabled the creation of a business use case diagram.

The process diagram in the business perspective was created based on two rules, the first of which is set in the partition element and has the form BauAp(v+)+, parameterized version: Ba(Client,Employee,Manager)uAp(Client)v(Send_request)p(Employee) v(Registration)p(Manager)v(Approve)p(Employee)v(Send_decision)p(Client)v(Receive). The second rule is set in the Subprocess element of the form RiviA(v+|i+)+, parameterized version: Ri(:Request)vi(:Decision)Av(Send_request)i([Sent]:Request) v(Registration)i([Registered:Request)v(Approve)i([Approved]:Request, [Approved]:Decision)v(Send_decision)i([Sent]:Decision)v(Receive) i([Received]:Decision]).

From the process diagram thus created, individual diagrams of system use cases, business class diagram and state machine diagram were generated. In the first case, the ApUau rule was set without parameters on the partition element named "Employee". In the second case, the AiviCczc rule was set on the action element named "4.Send_decision" (parameterized version: Ai(:Request)vi(:Decision) Ccz(1..*)c). In the third case, the AivSrst rule (without parameters) was used. This rule was set on the Instance element named "[Sent]:Decision" on the business use case realization diagram. In this way, a part of the software architecture was created in the context and business perspective.

The system use case realization diagram, in the system perspective, was created using three rules. The first one was set on the System use case in the form UauZ(pv+)+, parameterized form: UauZp(System)v(1.Form_display)p(User)v(2.Data_completion) p(System)v(3.Data_verification)p(User)v(4.Record_approval)p(System) v(5.Data_recording)p(User)v(6.End_of_action). The second rule was set on the class element named "Request" in the form CchZiv (parameterized version: CchZiv(1.Form_display)iv(3.Data_verification)iv(5.Data_recording)i ). In turn, the third rule was set on the Region element named "Request" in the form SrsZ(v+i+)+, parameterized form: SrsZi([Sent])v(1.Form_display)i([Filled])v(3.Data_verification) i([Verified])v(5.Data_recording)i([Registered]).

Based on the system use case realization diagram, an implementation use case diagram was created by setting the ZpvYuUu rule without parameters on the partition element named "System". In turn, the sequence diagram was created using the YuQ(lm+)+ rule on the implementation use case element named "Form display" (the parameterized version of this consistency rule is: Yu(Form_display) Ql(Registration) m(downLoad<Requests>::Request)l(:MQRabbit)m(downLoad<Request>::Request) l(Database) ). The last diagram in Figure 7 is a component diagram, which was created using the QlmlMqyq rule without parameters on the first message element of the sequence diagram named "downLoad(Requests)".

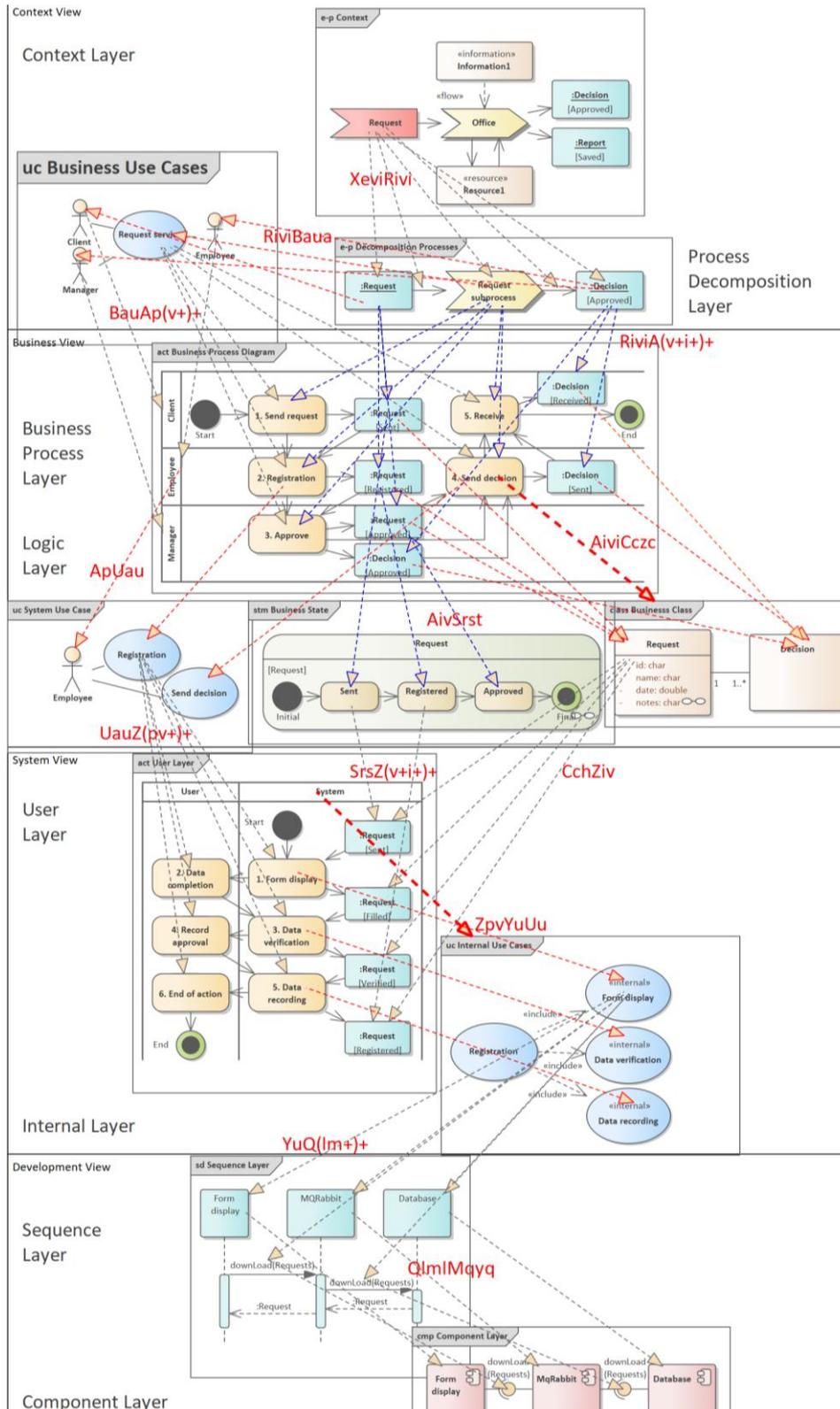

**Fig. 7.** An example of a software architecture fragment driven by composite consistency rules.

As shown in Figure 7, it is enough to create a context diagram and apply 13 composite consistency rules to obtain a software architecture consisting of 11 diagrams describing the system in four perspectives and in each of the two layers belonging to each perspective. These 13 rules replace 69 simple rules, that would make this figure unreadable and actually introduce some mess into the project. On the other hand, replacing the rules creating the business process, the system use case scenario, and the sequence diagram with appropriate composite consistency rules driven by patterns would further simplify the software architecture design process and drastically reduce its construction time.

## 5. Summary

To sum up the proposed composite consistency rules, it should be stated that these rules significantly simplify and speed up the construction of software architecture, because they propose a mechanism for creating certain aspects of software architecture, and not only enable mapping individual elements between individual diagrams, like the previous simple consistency rules.

The second aspect of optimal software architecture design is the possibility of using business process patterns, system use case realization diagram patterns, and sequence diagram patterns. Providing analysts, designers, or architects with the possibility of choosing a significant number of such patterns would probably optimize the construction of software architecture, or significantly shorten its design time, which would undoubtedly have an impact on shortening the time of building a given IT system.

Composite consistency rules are intuitive patterns that reflect the practical experience of software architects. They guide the systematic creation of consistent and complete software architectures using UML diagrams.

These rules are reusable across IT projects and help automate the design process, reducing errors and improving efficiency. By applying them, architects can speed up system development and gain greater confidence in the correctness and quality of the resulting architecture.

Some rules require manual intervention by the designer (it is the current state of any IT system design process), but pattern-driven composite consistency rules solve the issue perfectly.

The authors developed 29 composite consistency rules based on published simple ones. The list of them and their implementation as Enterprise Architect scripts are placed in folder JScript at [9].

The next step of our work is to develop a set of pattern-based consistency rules for business processes, system use case realization processes (Front-end processes), and sequence diagrams (Back-end processes) to further automate the generating of software architecture fragments and their models.